\documentclass[twocolumn,superscriptaddress,aps,prb,longbibliography]{revtex4-2}
\usepackage[utf8]{inputenc}
\usepackage[T1]{fontenc}
\usepackage{newtxtext,newtxmath}
\usepackage{graphicx}
\usepackage[colorlinks]{hyperref}
\usepackage{bm}
\usepackage{microtype}
\usepackage{physics}
\usepackage{xcolor}
\usepackage[caption=false]{subfig}
\usepackage{mwe}
\usepackage{relsize}

\begin{document}

\title{Emergent Chern metal in strongly correlated moiré bilayers}
\author{Kaushal Kumar Kesharpu}
\email{kesharpu@theor.jinr.ru}
\affiliation{Bogoliubov Laboratory of Theoretical Physics, Joint Institute for Nuclear Research, Dubna, Moscow Region 141980, Russia}

\author{Evgenii A. Kochetov}
\affiliation{Bogoliubov Laboratory of Theoretical Physics, Joint Institute for Nuclear Research, Dubna, Moscow Region 141980, Russia}

\date{\today}

\begin{abstract}
We investigate the topological properties of a strongly correlated moiré transition metal dichalcogenide (TMD) bilayer (such as twisted $\mathrm{MoTe}_2$ or $\mathrm{WSe}_2$) hosting a spin-charge separated state coupled to a layer-polarized helical spin texture. Analyzing an effective low-energy Hamiltonian for mobile spinless holons moving through this non-coplanar spinon background, we reveal pronounced asymmetric band warping. This warping closes the global indirect band gap across most of the parameter space, restricting true topological insulators to fine-tuned critical trajectories. However, a finite local direct gap persists, stabilizing broad regions of a topological Chern metal ($C = \pm 1$). Computing the Chern number and density of states, we demonstrate that this phase naturally yields a sizable, non-quantized anomalous Hall response accompanied by finite longitudinal conduction, providing a purely electronic strong-coupling mechanism for dissipative anomalous Hall states in moiré systems.
\end{abstract}

\maketitle

\section{Introduction}
\label{sec:introduction}

In low dimensions, quantum phenomena are significantly enhanced due to electronic confinement. Recently, moiré materials have emerged as a highly tunable platform for hosting exotic quantum phases. Specifically, in twisted moiré TMDs, the flatness of the emergent moiré bands---and consequently the strength of electronic correlations---can be continuously controlled via external electric field \cite{xu_InterplayTopologyCorrelations_NatPhys_2025,jia_MoireFractionalChern_PhysRevB_2024,li_ElectricallyTunedTopology_PhysRevB_2024}, lattice relaxation \cite{morales-duran_MagicAnglesFractional_PhysRevLett_2024,wang_FractionalChernInsulator_PhysRevLett_2024,mao_TransferLearningRelaxation_CommunPhys_2024} or twist angle \cite{deng_NonmonotonicBandFlattening_PhysRevX_2025,zhang_TwistangleTransferableContinuum_npjQuantumMater_2025,adak_TunableMoireMaterials_NatRevMater_2024}. This tunability provides an accessible regime for investigating strongly correlated phenomena \cite{li_QuantumPhasesTwisted_NatlSciRev_2026}, including the fractional quantum anomalous Hall effect \cite{li_QuantumAnomalousHall_Nature_2021}, Wigner crystallization \cite{kaushal_MagneticGroundStates_CommunPhys_2022,kuklin_ContrastingGKValley__2026}, and generalized Mott insulating phases \cite{li_ContinuousMottTransition_Nature_2021,devakul_MagicTwistedTransition_NatCommun_2021}.

Furthermore, twisted TMDs have emerged as a promising platform for hosting exotic spin-charge separation, a hallmark of fractionalized phases where spinless holons move through a background of spinful spinons. Specifically, parton mean-field approaches have predicted quantum phase transitions between superconducting and fractionalized insulating states \cite{kim_TheoryCorrelatedInsulators_NatCommun_2025}, while various chiral spin liquid phases have been widely explored in moiré bilayers \cite{zhang_SU4ChiralSpin_PhysRevLett_2021,crepel_SpinonPairingInduced_PhysRevLett_2024,motruk_KagomeChiralSpin_PhysRevResearch_2023,kuhlenkamp_ChiralPseudospinLiquids_PhysRevX_2024,zhou_QuantumPhasesTransition_PhysRevLett_2022,sun_SinglebandTriangularLattice_PhysRevB_2026,song_MottInsulatorsMoire_PhysRevB_2023,zhang_BridgingHubbardModel_PhysRevB_2019,zhang_DopingMottInsulator_PhysRevB_2022,delre_FieldControlSymmetrybroken_PhysRevRes_2024}. More recently, kinetic frustration has been identified as a robust mechanism to stabilize a spinon Fermi surface in these platforms \cite{kadow_FractionalizationKineticFrustration__2026}. Crucially, numerical evidence in twisted $\text{MoTe}_2$ shows that the spinless charge excitation gap can fall below its spinful counterpart, providing a distinct signature of spin-charge separation in moiré fractional Chern insulators \cite{goncalves_SpinlessSpinfulCharge_PhysRevLett_2026}. Experimentally, signature evidence of this separation has been reported via decoupled transport measurements, revealing diffusive spin transport alongside strongly quenched charge conduction in $\text{WSe}_2/\text{WS}_2$ heterobilayers \cite{regan_SpinTransportDoped_NatCommun_2024}. Beyond two dimensions, an emergent spin-charge separated Luttinger liquid phase has also been realized within anisotropic channels or one-dimensional domain walls of moiré TMD systems \cite{li_ImagingTunableLuttinger_Nature_2024,wu_PossibleSlidingRegimes_PhysRevLett_2024,zhang_EmergentOnedimensionalLuttinger_PhysRevB_2026}.

\begin{figure*}[tbh]
    \centering
    \includegraphics[width=\textwidth]{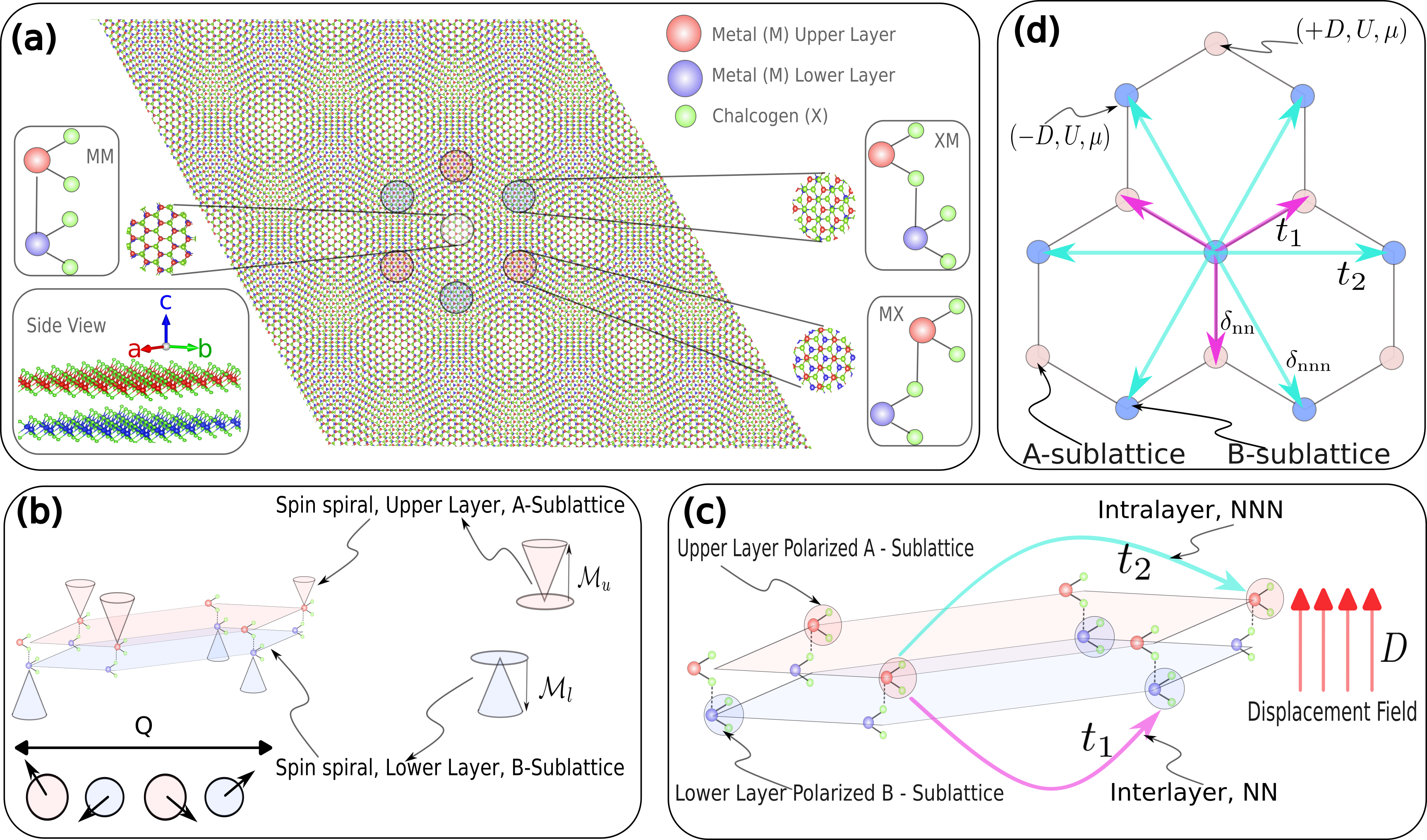}
    \caption{Schematic construction of the effective spin-charge separated model in a twisted transition metal dichalcogenide (TMD) moiré bilayer. 
    (a) Top and side views of the moiré superlattice, where red and blue spheres denote metal ($\mathrm{M}$) atoms in the upper and lower layers, respectively, and green spheres denote chalcogen ($\mathrm{X}$) atoms. Inequivalent stacking pockets ($\mathrm{MX}$ and $\mathrm{XM}$) localize low-energy Wannier orbitals predominantly on the upper ($A$) and lower ($B$) layers, forming an effective buckled honeycomb superlattice around the $\mathrm{MM}$ stacking regions. 
    (b) Layer-polarized conical spinon texture along the propagation wave vector $\mathbf{Q}$. In the strong-coupling Mott-Hubbard limit ($U \gg t_1, t_2$), local spins cant out of plane with opposite net magnetizations $+\mathcal{M}_u$ on sublattice $A$ (upper cones) and $-\mathcal{M}_l$ on sublattice $B$ (inverted lower cones) while rotating in plane with pitch $\mathbf{Q}$. 
    (c) Effective tight-binding parameters for spinless holons propagating in this non-coplanar background, illustrating interlayer nearest-neighbor (NN) hopping ($t_1$), intralayer next-nearest-neighbor (NNN) hopping ($t_2$), and the vertical electric displacement field ($D$) that tunes the sublattice potential offset. 
    (d) Top-down projection of the effective honeycomb lattice showing the three NN vectors $\boldsymbol{\delta}_{\text{nn}}$ (connecting $A$ and $B$ sublattices) and six NNN vectors $\boldsymbol{\delta}_{\text{nnn}}$ (within the same sublattice), along with sublattice dependent displacement field $\pm D$, Hubbard repulsion $U$ and chemical potential $\mu$.}
    \label{fig:moire-model}
\end{figure*}

A defining feature of moiré TMDs is the emergence of topological Chern insulator phases and the quantum anomalous Hall effect---along with their fractional counterparts---which can be tuned by twist angle, stacking order, or band occupancy \cite{cai_SignaturesFractionalQuantum_Nature_2023, zhao_RealizationHaldaneChern_NatPhys_2024}. Moreover, the unexpected Chern metal phase---where both chiral edge states and conducting bulk coexist---has been recently reported in moire TMDs \cite{zhao_EmergenceChernMetal_Science_2026,kumar_GatetunableHeavyFermion_PhysRevB_2022,guerci_ChiralKondoLattice_SciAdv_2023,crepel_AnomalousHallMetal_PhysRevB_2023,park_ObservationFractionallyQuantized_Nature_2023,park_ObservationDissipationlessFractional_NatPhys_2026,zhang_ThermoelectricityMoireHeavy_PhysRevLett_2026,han_TopologicalKondoInsulator_NatPhys_2026,guerci_TopologicalKondoSemimetal_PhysRevB_2024}. Microscopically, these topological phases are driven by emergent non-collinear or non-coplanar spin-valley textures, such as skyrmion lattices, where spin-up and spin-down densities localize on alternating sublattices of an effective honeycomb superlattice \cite{shi_AdiabaticApproximationAharonovCasher_PhysRevB_2024, thompson_MicroscopicSignaturesTopology_NatPhys_2025, zhang_SpintexturedChernBands_ProcNatlAcadSciUSA_2021, devakul_QuantumAnomalousHall_PhysRevX_2022}. Crucially, in some layer-polarized configurations, these sublattices correspond to physical top and bottom layers, mapping the spin-valley degrees of freedom directly onto a layer-locked pseudospin texture.

In a bilayer TMD system, an out-of-plane displacement field breaks the out-of-plane inversion symmetry and induces a layer-dependent Dzyaloshinskii-Moriya interaction (DMI) with a vector perpendicular to the monolayer planes (along the $z$-axis) \cite{pan_BandTopologyHubbard_PhysRevResearch_2020,rademaker_SpinorbitCouplingTransition_PhysRevB_2022,kim_TheoryCorrelatedInsulators_NatCommun_2025}. This interaction acts as a chiral torque that forces the local magnetic moments to rotate within the basal $xy$-plane. Concurrently, due to enhanced electronic correlations at integer or fractional fillings, the interlayer exchange interaction remains predominantly antiferromagnetic (AFM) \cite{xu_MaximallyLocalizedWannier_ProcNatlAcadSciUSA_2024, qiu_InteractionDrivenTopologicalPhase_PhysRevX_2023, chang_EvidenceCompetingGround_NatCommun_2026, crepel_BridgingSmallLarge_PhysRevResearch_2024, wang_BetheSlatercurvelikeBehaviorInterlayer_PhysRevB_2020}, favoring opposite out-of-plane spin alignment between the layers. In the presence of both competing mechanisms, the ground state compromises into a non-collinear, layer-polarized AFM spin-spiral structure characterized by opposite out-of-plane spin projections on the two layers ($+z$-axis on the upper layer and $-z$-axis on the lower layer).

Despite the physical relevance of this configuration, the interplay between layer-polarized AFM spin spiral texture and fractionalized electronic states remains largely unexplored. Specifically, we model the AFM spin textures on the effective honeycomb superlattice as shown in Fig.~\ref{fig:moire-model}(b). On the upper plane ($A$-sublattice), the spin profile is parameterized as:
\begin{equation}
  \label{eq:spiral-spin-upper}
  \begin{aligned}
    \mathbf{S}_{i,A} &= \left(S_{i,A}^{x}, S_{i,A}^{y}, S_{i,A}^{z}\right)\\
                     &= \frac{\sqrt{1 - \mathcal{M}_{u}^{2}}}{2} \left(\cos \mathbf{Q} \cdot \mathbf{r}_{i}, \, \sin \mathbf{Q} \cdot \mathbf{r}_{i}, \, \frac{\mathcal{M}_{u}}{\sqrt{1 - \mathcal{M}_{u}^{2}}} \right),
    \end{aligned}
\end{equation}
whereas on the lower plane ($B$-sublattice), the out-of-plane component is strictly inverted:
\begin{equation}
  \label{eq:spiral-spin-lower}
  \begin{aligned}
    \mathbf{S}_{i,B} &= \left(S_{i,B}^{x}, S_{i,B}^{y}, S_{i,B}^{z}\right)\\
                     &= \frac{\sqrt{1 - \mathcal{M}_{l}^{2}}}{2} \left(\cos \mathbf{Q} \cdot \mathbf{r}_{i}, \, \sin \mathbf{Q} \cdot \mathbf{r}_{i}, \, -\frac{\mathcal{M}_{l}}{\sqrt{1 - \mathcal{M}_{l}^{2}}} \right),
    \end{aligned}
\end{equation}
where $\mathbf{r}_{i}$ denotes the moiré lattice vector, and $\mathbf{Q}$ represents the in-plane propagation pitch vector of the helical modulation. The parameters $\mathcal{M}_{u}$ and $\mathcal{M}_{l}$ quantify the net out-of-plane magnetization components on the upper and lower planes, respectively. Crucially, both profiles preserve the strict constraint on the local spin amplitude, $|\mathbf{S}_{i,A/B}| = 1/2$.

In this work, we investigate the topological properties of a spin-charge separated state coupled to this layer-asymmetric magnetic background in twisted TMD bilayers. Employing the $su(2|1)$ coherent state formalism \cite{ferraz_EffectiveActionStrongly_NuclearPhysicsB_2011,kesharpu_MethodTreatStrongly_JPhysCondensMatter_2025,kesharpu_ProposalRealizingMajorana_PhysRevB_2024,kesharpu_ReentrantTopologicalOrder_PhysRevB_2025,kesharpu_TopologicalHallEffect_PhysRevB_2023,kesharpu_TopologicalHallEffect_PhysRevB_2024}, we construct an effective low-energy Hamiltonian for mobile spinless holons propagating through the AFM spin spiral spinon order. We show how the orientation of $\mathbf{Q}$ governs rotational symmetry breaking and valley warping, ultimately driving the system into a direct-gap-protected topological Chern metal over broad parameter regimes. Finally, by mapping the global phase diagram and evaluating the Chern number and density of states, we establish a purely strong-coupling mechanism for dissipative topological transport without invoking heavy-fermion Kondo hybridization \cite{kumar_GatetunableHeavyFermion_PhysRevB_2022,guerci_ChiralKondoLattice_SciAdv_2023}. The remainder of this paper is organized as follows: Sec.~\ref{sec:hamilt-disp} introduces the moiré-Hubbard model and the $su(2|1)$ fractionalization scheme; Sec.~\ref{sec:topol-prop} analyzes the topological band inversions and edge states, and also presents the global phase diagrams and density of states at Fermi level; Sec.~\ref{sec:discussion} discusses the conditions of applicability of the model, candidate materials and experimental probing techniques, along with summary of our findings.

\section{Hamiltonian and Dispersion}
\label{sec:hamilt-disp}

To model the electronic structure of the twisted TMD bilayer, we adopt an effective moiré-Hubbard Hamiltonian on a two-dimensional honeycomb superlattice \cite{wu_TopologicalInsulatorsTwisted_PhysRevLett_2019}. In this mapping, the $A$ and $B$ sublattices represent localized Wannier orbitals centered at high-symmetry stacking sites of the top and bottom layers, respectively \cite{wu_HubbardModelPhysicsTransition_PhysRevLett_2018}; also see Fig.~\ref{fig:moire-model}(a). Consequently, interlayer tunneling is described by the nearest-neighbor (NN) hopping $t_1$ between opposite sublattices ($\langle i,j \rangle$), while intralayer tunneling is captured by the next-nearest-neighbor (NNN) hopping $t_2$ within the same sublattice ($\langle\langle i,j \rangle\rangle$); also see Fig.~\ref{fig:moire-model}(c,d). An external vertical displacement field breaks the layer/sublattice degeneracy, introducing a staggered potential $\pm D$. Incorporating a local on-site Coulomb repulsion $U$ to account for strong electronic correlations, the full moiré-Hubbard Hamiltonian reads:
\begin{equation}
  \label{eq:ham-elec}
  \begin{aligned}
    H = & -t_1 \sum_{\langle i,j \rangle, \sigma} \left( a_{i,\sigma}^{\dagger} b_{j,\sigma} + \text{H.c.} \right) \\
        & -t_2 \sum_{\langle\langle i,j \rangle\rangle, \sigma} \left( a_{i,\sigma}^{\dagger} a_{j,\sigma} + b_{i,\sigma}^{\dagger} b_{j,\sigma} + \text{H.c.} \right) \\
        & + D \left( \sum_{i, \sigma} a_{i, \sigma}^{\dagger} a_{i, \sigma} - \sum_{j, \sigma} b_{j, \sigma}^{\dagger} b_{j, \sigma} \right) \\
        & + U \left( \sum_{i} n_{i,A,\uparrow} n_{i,A,\downarrow} + \sum_{j} n_{j,B,\uparrow} n_{j,B,\downarrow} \right) \\
        & - \mu \sum_{\sigma} \left( \sum_{i} a_{i, \sigma}^{\dagger} a_{i, \sigma} + \sum_{j} b_{j, \sigma}^{\dagger} b_{j, \sigma} \right),
  \end{aligned}
\end{equation}
where $a_{i,\sigma}^{\dagger}$ ($a_{i,\sigma}$) creates (annihilates) a fermion with spin $\sigma \in \{\uparrow, \downarrow\}$ on sublattice $A$ (upper layer) at site $i$, and $b_{j,\sigma}^{\dagger}$ ($b_{j,\sigma}$) acts analogously on sublattice $B$ (lower layer). The number operators are defined as $n_{i,A,\sigma} = a_{i,\sigma}^{\dagger} a_{i,\sigma}$ and $n_{j,B,\sigma} = b_{j,\sigma}^{\dagger} b_{j,\sigma}$, and $\mu$ is the chemical potential controlling the total carrier density.

In the strong correlation limit ($U \gg t_1, t_2$), doubly occupied states are projected out, restricting the low-energy Hilbert space to singly occupied and empty sites. To treat the strong correlation limit without introducing unphysical doubly occupied states, we employ the $su(2|1)$ coherent state representation \cite{perelomov_GeneralizedCoherentStates__1986, ferraz_EffectiveActionStrongly_NuclearPhysicsB_2011,kesharpu_MethodTreatStrongly_JPhysCondensMatter_2025,kesharpu_ProposalRealizingMajorana_PhysRevB_2024,kesharpu_ReentrantTopologicalOrder_PhysRevB_2025,kesharpu_TopologicalHallEffect_PhysRevB_2023,kesharpu_TopologicalHallEffect_PhysRevB_2024}. The local state on sublattice $\alpha \in \{A, B\}$ is parameterized as:
\begin{equation}
  \label{eq:defn-coherent-state-sublattice}
  \ket{z_{i,\alpha}, \xi_{i,\alpha}} = \frac{\ket{\uparrow}_{\alpha} + z_{i,\alpha} \ket{\downarrow}_{\alpha} + \xi_{i,\alpha} \ket{0}_{\alpha}}{\sqrt{1 + z_{i,\alpha} \bar{z}_{i,\alpha}} + \xi_{i, \alpha} \bar{\xi}_{i, \alpha}},
\end{equation}
where $z_{i,\alpha}$ is a complex field representing the localized spinon degree of freedom, and $\xi_{i,\alpha}$ is an odd Grassmann field representing the mobile, spinless holon. Unlike slave-particle frameworks, the $su(2|1)$ algebra automatically prevents double occupancy without requiring artificial gauge fields or explicit local constraints.

The local spin components map onto the spinon fields via $z_{i} = [2 / (1 + 2 S_{i}^{z})] \sqrt{(S_{i}^{x})^{2} + (S_{i}^{y})^{2}} \exp [ \imath \arctan (S_{i}^{y}/S_{i}^{x}) ]$. Substituting our layer-polarized AFM spin spiral configurations [Eqs.~\eqref{eq:spiral-spin-upper} and \eqref{eq:spiral-spin-lower}] into this relation yields the real-space spinon profiles:
\begin{equation}
  \label{eq:zi-fields-opposite}
  z_{i,A} = \sqrt{\frac{1-\mathcal{M}_{u}}{1+\mathcal{M}_{u}}} \mathrm{e}^{\imath \mathbf{Q} \cdot \mathbf{r}_{i}}, \quad 
  z_{j,B} = \sqrt{\frac{1+\mathcal{M}_{l}}{1-\mathcal{M}_{l}}} \mathrm{e}^{\imath \mathbf{Q} \cdot \mathbf{r}_{j}}.
\end{equation}
By evaluating the kinetic terms of the moiré-Hubbard model under this coherent state background, the effective mean-field Hamiltonian for the holons in momentum space takes the form (see Appendix for details):
\begin{equation}
  \label{eq:H_k_defn-doped}
  \mathcal{H}_{\mathbf{k}} =
  \begin{bmatrix}
    t_2 \gamma_{\mathbf{k}, nnn}^{A} - \mu + D & t_1 \gamma_{\mathbf{k}, nn}^{\ast}\\
    t_1 \gamma_{\mathbf{k}, nn} & t_2 \gamma_{\mathbf{k}, nnn}^{B} - \mu - D
  \end{bmatrix}.
\end{equation}
Defining $\mathcal{M}_{u-l} = (1 + \mathcal{M}_{u}) (1 - \mathcal{M}_{l})$ and $\mathcal{M}_{l-u} = (1 + \mathcal{M}_{l}) (1 - \mathcal{M}_{u})$, the NN structural factor becomes:
\begin{equation}
  \label{eq:defn-gamma-k-nn}
  \gamma_{\mathbf{k},nn} = \sum_{\boldsymbol{\delta}_{n}} \left( \sqrt{\mathcal{M}_{u-l}}\mathrm{e}^{-\imath \frac{\mathbf{Q} \cdot \boldsymbol{\delta}_{nn}}{2}} + \sqrt{\mathcal{M}_{l-u}} \mathrm{e}^{\imath \frac{\mathbf{Q} \cdot \boldsymbol{\delta}_{nn}}{2}} \right) \mathrm{e}^{-\imath \mathbf{k} \cdot \boldsymbol{\delta}_{nn}},
\end{equation}
Similarly, the NNN structural factors within each layer are given by:
\begin{equation}
  \label{eq:defn-gamma-k-nnn}
  \begin{aligned}
    \gamma_{\mathbf{k}, nnn}^{A} &= \sum_{\boldsymbol{\delta}_{nnn}} \left( \cos \frac{\mathbf{Q} \cdot \boldsymbol{\delta}_{nnn}}{2} - \imath \mathcal{M}_{u} \sin \frac{\mathbf{Q} \cdot \boldsymbol{\delta}_{nnn}}{2} \right) \mathrm{e}^{-\imath \mathbf{k} \cdot \boldsymbol{\delta}_{nnn}},\\
    \gamma_{\mathbf{k}, nnn}^{B} &= \sum_{\boldsymbol{\delta}_{nnn}} \left( \cos \frac{\mathbf{Q} \cdot \boldsymbol{\delta}_{nnn}}{2} + \imath \mathcal{M}_{l} \sin \frac{\mathbf{Q} \cdot \boldsymbol{\delta}_{nnn}}{2} \right) \mathrm{e}^{-\imath \mathbf{k} \cdot \boldsymbol{\delta}_{nnn}},
  \end{aligned}
\end{equation}
Notably, the overall signs of the effective hopping terms in Eq.~\eqref{eq:H_k_defn-doped} are inverted relative to the bare electronic parameters. This is a robust feature arising directly from the anti-commutation relations of the odd Grassmann holon fields during real-space transport.

\begin{figure*}[tbh]
    \centering
    \includegraphics[width=\textwidth]{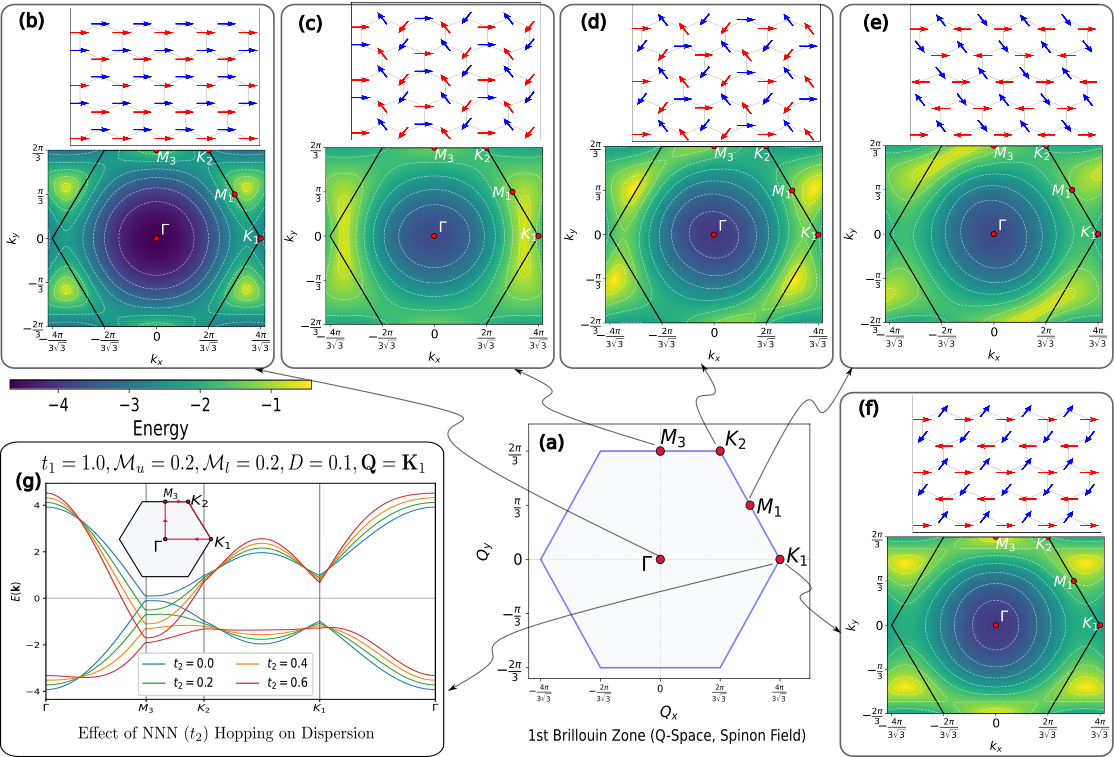}
    \caption{Evolution of the spinon magnetic texture and effective holon band dispersion across the moiré Brillouin zone. 
    (a) First Brillouin zone in $\mathbf{Q}$-space highlighting five representative spiral pitch vectors: $\mathbf{\Gamma}$, $\mathbf{M}_3$, $\mathbf{K}_2$, $\mathbf{M}_1$, and $\mathbf{K}_1$. 
    (b)--(f) Corresponding real-space in-plane spinon configurations on the $A$ (red arrows) and $B$ (blue arrows) sublattices (top subpanels) alongside momentum-space intensity maps of the lower holon band $E_{-}(\mathbf{k})$ (bottom subpanels, with the moiré Brillouin zone outlined in black). Starting from the rotationally symmetric collinear state at $\mathbf{Q} = \mathbf{\Gamma}$ (b), finite pitch vectors reshape the dispersion depending on their alignment with crystal mirror axes: axis-aligned stripe ($\mathbf{M}_3$, c) and $120^\circ$ spiral ($\mathbf{K}_1$, f) textures preserve reflection symmetry and deform the energy contours symmetrically, whereas off-axis textures ($\mathbf{K}_2$, d; $\mathbf{M}_1$, e) break mirror planes and shear the valleys diagonally. 
    (g) Band structure $E_{\pm}(\mathbf{k})$ along the high-symmetry contour $\mathbf{\Gamma}\text{--}\mathbf{M}_3\text{--}\mathbf{K}_2\text{--}\mathbf{K}_1\text{--}\mathbf{\Gamma}$ (inset) at $\mathbf{Q} = \mathbf{K}_1$ for $t_1 = 1.0$, $\mathcal{M}_u = \mathcal{M}_l = 0.2$, and varying intralayer NNN hopping $t_2 \in \{0.0, 0.2, 0.4, 0.6\}$. Solid and dashed lines denote the upper ($E_{+}$) and lower ($E_{-}$) holon bands, respectively, demonstrating how increasing $t_2$ flattens the valence band $E_{-}(\mathbf{k})$ and drives indirect band overlap.}
    \label{fig:dispersion-Q-t2}
\end{figure*}

Diagonalizing Eq. (\ref{eq:H_k_defn-doped}) yields two energy bands:
\begin{equation}
  \label{eq:energ-disp}
  \begin{aligned}
    E_{\pm}(\mathbf{k}) =&  \left[ t_{2} \left( \frac{\gamma_{\mathbf{k},nnn}^{A} + \gamma_{\mathbf{k},nnn}^{B}}{2} \right) - \mu  \right]\\
                         &\quad \pm \sqrt{\left[ D + t_{2} \left( \frac{\gamma_{\mathbf{k},nnn}^{A} - \gamma_{\mathbf{k},nnn}^{B}}{2}  \right) \right]^{2} + t_{1}^2 \left| \gamma_{\mathbf{k}, nn} \right|^2}.
  \end{aligned}
\end{equation}
In Eq.~(\ref{eq:energ-disp}), the first term acts as a momentum-dependent diagonal energy shift that warps the Fermi surface. Concurrently, the energy gap
\begin{equation}
  \label{eq:gap-expa}
  \begin{aligned}
    &\Delta(\mathbf{k}) \\
    &= 2 \sqrt{\left[ D - t_{2} \left( \frac{\mathcal{M}_{u} + \mathcal{M}_{l}}{2} \right) \sin  \frac{ \mathbf{Q} \cdot \boldsymbol{\delta}_{nnn} }{2} \sin \mathbf{k} \cdot \boldsymbol{\delta}_{nnn} \right]^{2} + t_{1}^2 \left| \gamma_{\mathbf{k}, nn} \right|^2},
  \end{aligned}
\end{equation}
which is an emergent topological mass generated by the spinon background is dynamically governed by external displacement field $D$, layer magnetization $\mathcal{M}_{u}$ and $\mathcal{M}_{l}$, NNN hopping $t_{2}$ and spiral modulation vector $\mathbf{Q}$.

We illustrate how the spatial spin modulation reshapes the holon kinetics. Fig.~\ref{fig:dispersion-Q-t2} shows the lower band dispersion $E_{-}(\mathbf{k})$ across the moiré Brillouin zone for five representative pitch vectors $\mathbf{Q}$. At $\mathbf{Q} = \mathbf{\Gamma}$, the spin background is collinear and preserves all lattice symmetries, yielding a deep central minimum at $\mathbf{\Gamma}$ surrounded by four symmetric peaks. Introducing a finite spiral pitch $\mathbf{Q}$ partially frustrates the hopping, which raises the central minimum and narrows the overall bandwidth. Crucially, the shape of the resulting band warping depends on the orientation of $\mathbf{Q}$ relative to the crystal mirror planes. When $\mathbf{Q}$ aligns with the orthogonal mirror axes of the superlattice—specifically at $\mathbf{Q} = \mathbf{K}_1$ (along $k_x$) and $\mathbf{Q} = \mathbf{M}_3$ (along $k_y$)—reflection symmetry is preserved, and the dispersion deforms symmetrically into straight vertical ridges. In contrast, choosing $\mathbf{Q} = \mathbf{K}_2$ or $\mathbf{Q} = \mathbf{M}_1$ breaks these mirror planes, causing the spiral background to shear and tilt the energy contours diagonally. This directional control over valley tilting and bandwidth provides the essential electronic asymmetry needed to generate non-trivial Berry curvature.

\section{Topological Properties}
\label{sec:topol-prop}

\begin{figure*}[tbh]
    \centering
    \includegraphics[width=\textwidth]{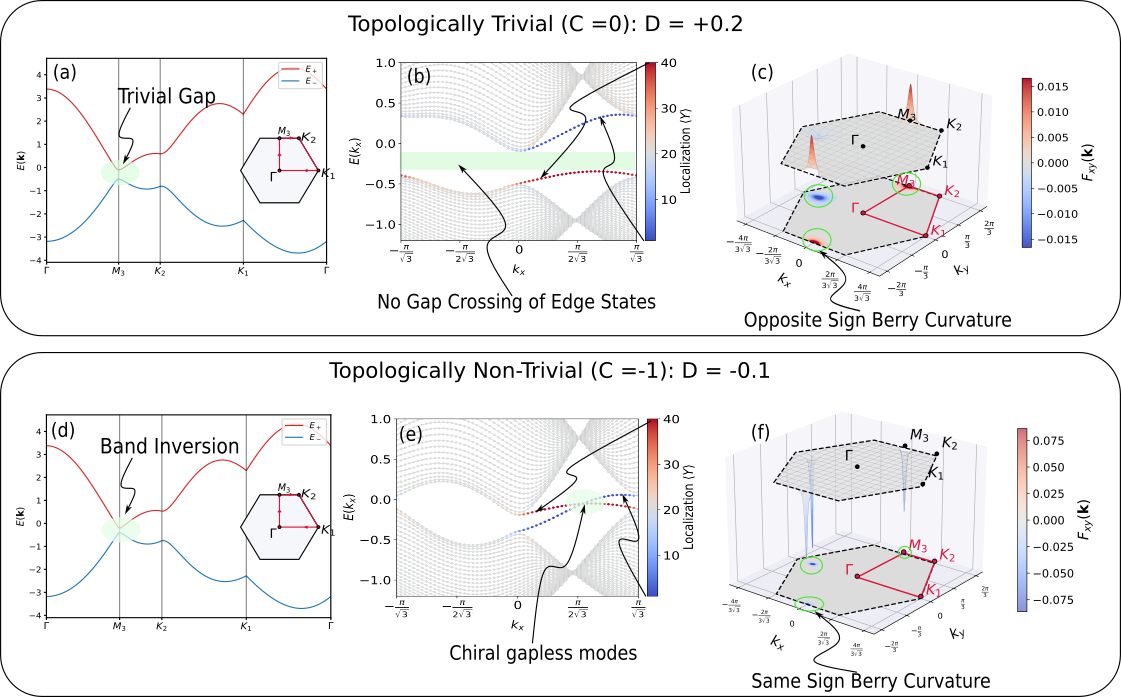}
    \caption{Topological phase transition driven by the vertical displacement field $D$, contrasting the topologically trivial state at $D = +0.2$ ($C = 0$, top row) with the non-trivial topological state at $D = -0.1$ ($C = -1$, bottom row) for fixed $t_1 = 1.0$, $t_2 = 0.2$, $\mu = 0.0$, $\mathcal{M}_u = 0.9$, $\mathcal{M}_l = 0.2$, and $\mathbf{Q} = \mathbf{K}_1$. 
    (a), (d) Bulk holon band structures $E_{\pm}(\mathbf{k})$ along the high-symmetry path $\mathbf{\Gamma}\text{--}\mathbf{M}_3\text{--}\mathbf{K}_2\text{--}\mathbf{K}_1\text{--}\mathbf{\Gamma}$ (insets). Tuning $D$ from $+0.2$ to $-0.1$ closes and reopens the local direct gap at the $\mathbf{M}_3$ point (green circles), inducing a topological band inversion. 
    (b), (e) Energy spectra $E(k_x)$ of a finite ribbon of width $N_y = 40$, where the color scale indicates the average transverse position $\langle Y \rangle$ of each eigenstate (red and blue denote states localized on opposite edges, while gray denotes extended bulk states). In the trivial phase (b), edge states remain detached within the bulk gap (green shaded region) without connecting the valence and conduction bands. Following band inversion (e), counter-propagating chiral edge modes bridge the local direct gap across the opposite boundaries while coexisting in energy with bulk states elsewhere in the Brillouin zone, realizing a direct-gap-protected Chern metal. 
    (c), (f) Momentum-space distribution of the lattice Berry curvature $F_{xy}(\mathbf{k})$ across the first moiré Brillouin zone (dashed upper hexagon and lower 2D projection). In the trivial phase (c), positive (red) and negative (blue) Berry curvature peaks at the three inequivalent BZ points (green circles) carry opposite signs and cancel out ($C = 0$). In the topological phase (f), all three points develop sharp negative peaks of the same sign, integrating to a quantized Chern number $C = -1$.}
    \label{fig:topology-transition}
\end{figure*}

To characterize the topological properties of the system, it is convenient to project the momentum-space Hamiltonian $\mathcal{H}_{\mathbf{k}}$ onto the basis of Pauli matrices:
\begin{equation}
  \label{eq:Hk-pseudo-spin}
  \mathcal{H}_{\mathbf{k}} = d_{0}(\mathbf{k}) \sigma_{0} + \mathbf{d}(\mathbf{k}) \cdot \boldsymbol{\sigma},
\end{equation}
where $\sigma_0$ is the $2 \times 2$ identity matrix and $\boldsymbol{\sigma} = (\sigma_x, \sigma_y, \sigma_z)$ denotes the vector of Pauli matrices acting on the layer pseudospin space. The components of the $\mathbf{d}(\mathbf{k})$ vector are explicitly given by:
\begin{equation*}
  \label{eq:defn-d-k}
  \begin{aligned}
    d_{0}(\mathbf{k}) &= t_2 \left( \frac{\gamma_{\mathbf{k},nnn}^{A} + \gamma_{\mathbf{k},nnn}^{B}}{2} \right) - \mu, 
    &d_{x}(\mathbf{k}) &= t_1 \Re \left[ \gamma_{\mathbf{k},nn} \right],\\
    d_{z}(\mathbf{k}) &= t_2 \left( \frac{\gamma_{\mathbf{k},nnn}^{A} - \gamma_{\mathbf{k},nnn}^{B}}{2} \right) + D,
    &d_{y}(\mathbf{k}) &= -t_1 \Im \left[ \gamma_{\mathbf{k},nn} \right].
  \end{aligned}
\end{equation*}
Crucially, because the effective mass term $d_{z}(\mathbf{k})$ is not a rigid constant but varies continuously across the Brillouin zone due to the emergent Haldane-like kinetic contribution, the vector $\mathbf{d}(\mathbf{k})$ maps the torus of the Brillouin zone onto a closed, non-planar surface in parameter space rather than a flat sheet. This geometric winding provides the non-zero solid angle required to sweep out a non-trivial Chern number.

The Chern number for the occupied lower band is calculated by integrating the Berry curvature over the first Brillouin zone:
\begin{equation}
  \label{eq:chern-number-2band} 
  C = \frac{1}{4 \pi} \int_{\text{BZ}} \mathrm{d}^2k \, \frac{1}{|\mathbf{d}(\mathbf{k})|^3} \, \mathbf{d}(\mathbf{k}) \cdot \left( \frac{\partial \mathbf{d}(\mathbf{k})}{\partial k_x} \times \frac{\partial \mathbf{d}(\mathbf{k})}{\partial k_y}\right).
\end{equation}
To demonstrate the topological phase transition driven by the external layer-polarizing field, Fig.~\ref{fig:topology-transition} contrasts the bulk band structures, finite-ribbon spectra, and Berry curvature distributions for opposite signs of the vertical displacement field $D$. At $D = +0.2$ [Figs.~\ref{fig:topology-transition}(a)--\ref{fig:topology-transition}(c)], the system resides in a topologically trivial insulating state ($C = 0$). Here, the bulk bands are separated by a clean global gap around $\mathbf{M}_3$ [Fig.~\ref{fig:topology-transition}(a)], while the ribbon spectrum [Fig.~\ref{fig:topology-transition}(b)] hosts detached boundary modes within the gap that fail to connect the valence and conduction bands. Correspondingly, the Berry curvature $F_{xy}(\mathbf{k})$ exhibits peaks of opposite signs across the three inequivalent BZ points [Fig.~\ref{fig:topology-transition}(c)], canceling out upon integration over the Brillouin zone ($C = 0$).

In sharp contrast, reversing the displacement field to $D = -0.1$ [Figs.~\ref{fig:topology-transition}(d)--\ref{fig:topology-transition}(f)] drives the system across a topological phase transition via a band inversion at the high-symmetry $\mathbf{M}_3$ point, closing and reopening the local direct gap to yield $C = -1$. In the ribbon geometry [Fig.~\ref{fig:topology-transition}(e)], this inversion produces counter-propagating chiral edge modes that bridge the local direct gap on opposite boundaries, accompanied by negative Berry curvature peaks of the same sign across all three BZ points [Fig.~\ref{fig:topology-transition}(f)]. Crucially, because this band inversion occurs on top of a heavily warped dispersion background, the chiral edge modes coexist in energy with bulk states elsewhere in the Brillouin zone, realizing a direct-gap-protected topological Chern metal. Tuning the vertical displacement field thus provides a direct experimental knob to trigger band inversion and switch on protected boundary transport.

\begin{figure*}[tbh]
    \centering
    \includegraphics[width=\textwidth]{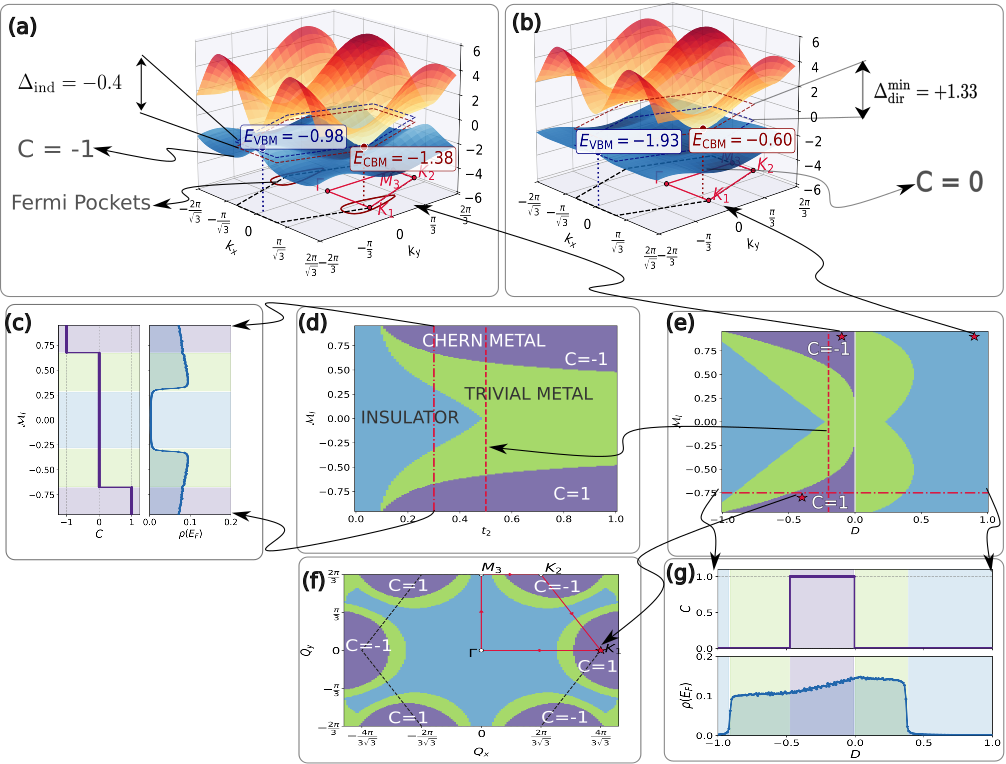}
    \caption{Global topological phase diagrams, representative three-dimensional holon dispersions, and density-of-states profiles of the spin-charge separated moiré bilayer. 
    (a), (b) Three-dimensional energy dispersions $E_{\pm}(\mathbf{k})$ across the first moiré Brillouin zone (red hexagon) evaluated at two representative parameter points (red stars in panel (e)). In the Chern metal regime (a) at $(D, \mathcal{M}_l) = (-0.1, 0.9)$, severe band warping pushes the valence-band maximum ($E_{\text{VBM}} = -0.98$) above the conduction-band minimum ($E_{\text{CBM}} = -1.38$), yielding a negative indirect gap ($\Delta_{\text{ind}} = -0.4$) with bulk Fermi pockets while preserving a finite local direct gap and a quantized Chern number $C = -1$. In the trivial insulator regime (b) at $(D, \mathcal{M}_l) = (0.9, 0.9)$, a large positive minimum of direct gap over whole BZ ($\Delta_{\text{dir}}^{\text{min}} = +1.33$) separates the bands ($E_{\text{VBM}} = -1.93$, $E_{\text{CBM}} = -0.60$) with $C = 0$. 
    (c) Vertical line cut along $t_2 = 0.3$ (red dash-dotted line in panel (d)), displaying the Chern number $C$ (left) and the density of states at the Fermi level $\rho(E_F)$ (right) as a function of $\mathcal{M}_l$. 
    (d) Phase diagram in the $(t_2, \mathcal{M}_l)$ plane at $D = -0.2$ (vertical red dashed line in panel (e)), showing symmetric $C = -1$ ($\mathcal{M}_l > 0$) and $C = +1$ ($\mathcal{M}_l < 0$) Chern metal lobes (purple) alongside the trivial indirect metal (green) and trivial band insulator (blue) phases. 
    (e) Global phase diagram in the $(D, \mathcal{M}_l)$ plane for fixed $\mathcal{M}_u = 0.0$, $t_1 = 1.0$, $t_2 = 0.3$, and $\mathbf{Q} = \mathbf{K}_1$. 
    (f) Phase evolution across the first Brillouin zone in $\mathbf{Q}$-space (black dashed hexagon) evaluated at $(D, \mathcal{M}_l) = (-0.4, -0.8)$ (red star in panel (e)), illustrating how incommensurate spiral modulations stabilize alternating $C = \pm 1$ Chern metal regions. 
    (g) Horizontal line cut along $\mathcal{M}_l = -0.75$ (red dash-dotted line in panel (e)), showing the quantized $C = +1$ plateau (top) coexisting with a finite Fermi-level density of states $\rho(E_F) > 0$ (bottom) across the Chern metal regime before $\rho(E_F)$ vanishes upon entering the trivial insulator at larger $D$.}
    \label{fig:global-phase-diagram}
\end{figure*}

A comprehensive understanding of the system requires mapping out the global stability of these phases across the parameter space. To classify the emergent states systematically, we evaluate the minimum local direct band gap,
$$\Delta_{\text{dir}}^{\min} = \min_{\mathbf{k}} [E_{+}(\mathbf{k}) - E_{-}(\mathbf{k})],$$
alongside the global indirect gap,
$$\Delta_{\text{ind}} = \min_{\mathbf{k}}[E_{+}(\mathbf{k})] - \max_{\mathbf{k}}[E_{-}(\mathbf{k})].$$
Parameter regions where $\Delta_{\text{dir}}^{\min} \le 0$ correspond to gapless metals where the Chern number is ill-defined. When a finite direct gap is preserved ($\Delta_{\text{dir}}^{\min} > 0$), a positive indirect gap ($\Delta_{\text{ind}} > 0$) defines an insulating state—classified as a true topological insulator ($C \neq 0$) or a trivial band insulator ($C = 0$). Conversely, a negative indirect gap ($\Delta_{\text{ind}} \le 0$) signals an indirect band-overlap metal, separating into a direct-gap-protected topological metal ($C \neq 0$) or a trivial metal ($C = 0$).

Fig.~\ref{fig:global-phase-diagram} presents the global phase diagrams alongside representative 3D holon dispersions and density-of-states (DOS) profiles. Figure~\ref{fig:global-phase-diagram}(e) shows the phase diagram in the $(D, \mathcal{M}_l)$ plane for fixed upper-layer magnetization $\mathcal{M}_u = 0$ and spiral modulation $\mathbf{Q} = \mathbf{K}_1$. In this regime, Chern metal phases ($C = \pm 1$) emerge asymmetrically for $D < 0$ alongside broad trivial metal regions, reflecting the direct competition between the external displacement field and the chiral spinon mass; tuning $\mathcal{M}_u$ within $-1 < \mathcal{M}_u < 1$ simply introduces an internal layer-bias offset that shifts the phase boundaries vertically. Figure~\ref{fig:global-phase-diagram}(d) illustrates the impact of the NNN hopping $t_2$ in the $(t_2, \mathcal{M}_l)$ plane along the cut $D = -0.2$, where increasing $t_2$ simultaneously amplifies band warping and the Haldane-like mass term $\propto t_2(\mathcal{M}_u + \mathcal{M}_l)$ in Eq.~\eqref{eq:gap-expa}, yielding phase boundaries symmetric about $\mathcal{M}_l = 0$. Finally, Fig.~\ref{fig:global-phase-diagram}(f) maps the phase evolution across the full BZ of the spiral pitch vector $\mathbf{Q}$. Long-wavelength modulations near $\mathbf{\Gamma}$ remain topologically trivial due to insufficient spatial spin twist, whereas moving outward toward the zone boundaries stabilizes the Chern metal phase.

A striking feature of the global phase space is that true topological insulators are restricted to narrow, fine-tuned critical trajectories. Because the non-collinear AFM spiral modulation and intralayer hopping $t_2$ severely warp the holon dispersion, the global indirect gap closes well before the local direct gap is destroyed. As illustrated by the 3D dispersions in Figs.~\ref{fig:global-phase-diagram}(a) and \ref{fig:global-phase-diagram}(b), in the Chern metal regime at $(D, \mathcal{M}_l) = (-0.1, 0.9)$ [Fig.~\ref{fig:global-phase-diagram}(a)], the valence-band maximum ($E_{\text{VBM}} = -0.98$) rises above the conduction-band minimum ($E_{\text{CBM}} = -1.38$), yielding a negative indirect gap ($\Delta_{\text{ind}} = -0.4$) with bulk Fermi pockets while preserving a finite local direct gap and $C = -1$. In contrast, at $(D, \mathcal{M}_l) = (0.9, 0.9)$ [Fig.~\ref{fig:global-phase-diagram}(b)], a large positive gap ($\Delta_{\text{dir}}^{\text{min}} = +1.33$) opens across the entire BZ, stabilizing a trivial band insulator ($C = 0$).

To confirm the coexistence of metallic bulk carriers with non-trivial band topology, we evaluate the density of states at the Fermi level,
$$\rho(E_F) = \sum_{n=\pm} \int_{\text{BZ}} \frac{\mathrm{d}^2k}{(2\pi)^2} \: \text{Im} \left[ \frac{1}{E_{F} - E_{n}(\mathbf{k}) + \imath \eta} \right] ,$$
along representative line cuts across the phase boundaries (here $\eta$ small is the broadening parameter). In the vertical cut along $t_2 = 0.3$ [Fig.~\ref{fig:global-phase-diagram}(c)] and the horizontal cut along $\mathcal{M}_l = -0.75$ [Fig.~\ref{fig:global-phase-diagram}(g)], $\rho(E_F)$ vanishes inside the trivial insulating gap ($C = 0$, blue regions) but remains strictly finite ($\rho(E_F) > 0$) across the entire $C = \pm 1$ Chern metal plateaus (purple regions). Because these partially occupied bulk Fermi pockets carry non-zero Berry curvature, they prevent exact quantization of the intrinsic anomalous Hall conductivity $\sigma_{xy}$ while providing dissipative conduction channels—naturally accounting for a sizable anomalous Hall signal accompanied by finite longitudinal resistance ($R_{xx} \neq 0$) in moiré TMD transport experiments.

\section{Discussion and Conclusion}
\label{sec:discussion}

We would like to delineate the physical regime of applicability and key approximations of our theoretical framework. First, the $su(2|1)$ coherent-state representation strictly projects out doubly occupied sites, confining our model to the strong-coupling Mott-Hubbard limit ($U \gg t_1, t_2$) near single-hole filling ($\nu = 1 \pm \delta$) realized in small-angle moiré TMDs \cite{wu_HubbardModelPhysicsTransition_PhysRevLett_2018,tang_SimulationHubbardModel_Nature_2020,regan_MottGeneralizedWigner_Nature_2020,xu_CorrelatedInsulatingStates_Nature_2020,mak_SimulatingHubbardModel_NatlSciRev_2026}. Second, by parameterizing the spinon sector as a static, layer-polarized AFM spiral, our effective holon Hamiltonian applies at low temperatures ($T \ll J$) where long-range or quasi-static non-collinear magnetic order is established, while neglecting dynamical $U(1)$ gauge fluctuations that may renormalize the metallic phase boundaries near continuous transitions \cite{pan_BandTopologyHubbard_PhysRevResearch_2020,qiu_InteractionDrivenTopologicalPhase_PhysRevX_2023,he_ItinerantMagnetismTriangularlattice_PhysRevB_2026,hejazi_NoncollinearPhasesMoire_ProcNatlAcadSciUSA_2020,chang_EvidenceCompetingGround_NatCommun_2026,lee_DopingMottInsulator_RevModPhys_2006,tao_ValleyCoherentQuantumAnomalous_PhysRevX_2024,kaushal_MagneticGroundStates_CommunPhys_2022}. Third, restricting the tight-binding basis to the lowest two layer-polarized Wannier orbitals remains valid provided the vertical displacement field $D$ and kinetic bandwidth remain smaller than the energy separation to higher remote moiré minibands \cite{wu_TopologicalInsulatorsTwisted_PhysRevLett_2019,devakul_MagicTwistedTransition_NatCommun_2021,wang_FractionalChernInsulator_PhysRevLett_2024,zhang_PolarizationdrivenBandTopology_NatCommun_2024}. Within these well-defined boundaries, our model provides a controlled, minimal description of how non-coplanar spin textures drive fractionalized Chern metal states.

It is also important to situate our single-band fractionalized framework alongside moiré Kondo and chiral Kondo lattice theories developed for twisted TMD heterobilayers \cite{kumar_GatetunableHeavyFermion_PhysRevB_2022,guerci_ChiralKondoLattice_SciAdv_2023}. In the Kondo lattice picture, a large intrinsic band offset or strong displacement field locks one layer at integer Mott filling to form localized magnetic moments, while doped carriers occupy a separate, weakly correlated itinerant band that hybridizes with those moments via Kondo exchange. By contrast, our $su(2|1)$ formulation describes the regime of strong interlayer hybridization—such as in twisted homobilayers or heterobilayers tuned near layer-degeneracy—where low-energy holes reside within a single strongly correlated honeycomb Hubbard band ($U \gg t_1, t_2$) doped away from half-filling ($\nu = 1 \pm \delta$) \cite{wu_HubbardModelPhysicsTransition_PhysRevLett_2018,tang_SimulationHubbardModel_Nature_2020,regan_MottGeneralizedWigner_Nature_2020,xu_CorrelatedInsulatingStates_Nature_2020,mak_SimulatingHubbardModel_NatlSciRev_2026}. In the absence of a separate itinerant spectator band, the doped carriers themselves undergo spin-charge separation \cite{andrei_UniversalMagneticEnergy__2026,kadow_FractionalizationKineticFrustration__2026}, and non-trivial Berry curvature emerges intrinsically from holons propagating through the non-coplanar spinon spiral \cite{ohgushi_SpinAnisotropyQuantum_PhysRevB_2000}. Rather than competing, the two mechanisms represent complementary limits of strongly correlated moiré TMDs: while Kondo hybridization governs the strongly layer-asymmetric two-band regime \cite{kumar_GatetunableHeavyFermion_PhysRevB_2022,guerci_ChiralKondoLattice_SciAdv_2023}, our spin-charge separated Chern metal captures dissipative topological transport when both layers actively participate in correlated hopping.

Our theoretical predictions offer several direct signatures for transport and optical experiments in dual-gated moiré TMD bilayers, such as twisted $\mathrm{MoTe}_2$ and $\mathrm{WSe}_2$ \cite{anderson_ProgrammingCorrelatedMagnetic_Science_2023,cai_SignaturesFractionalQuantum_Nature_2023,zeng_ThermodynamicEvidenceFractional_Nature_2023}. First, tuning the vertical displacement field $D$ across the topological phase boundary should drive a transition from a trivial insulator/metal into the Chern metal regime, marked by the onset of a sizable, non-quantized anomalous Hall resistance ($R_{xy} \neq 0$) accompanied by a finite longitudinal resistance ($R_{xx} \neq 0$) due to the warped bulk Fermi pockets. Second, because unequal layer magnetizations ($\mathcal{M}_u \neq \mathcal{M}_l$) shift the phase boundaries vertically in the $(D, \mathcal{M}_l)$ plane, the onset of the anomalous Hall signal is inherently asymmetric with respect to the sign of $D$, while out-of-plane magnetic-field reverses the sign of $R_{xy}$ via magnetic hysteresis \cite{anderson_ProgrammingCorrelatedMagnetic_Science_2023,li_QuantumAnomalousHall_Nature_2021,foutty_MappingTwisttunedMultiband_Science_2024}. Finally, combining transport with reflective magnetic circular dichroism (RMCD) \cite{anderson_ProgrammingCorrelatedMagnetic_Science_2023,cai_SignaturesFractionalQuantum_Nature_2023} and local magnetic imaging—such as nano-SQUID magnetometry \cite{redekop_DirectMagneticImaging_Nature_2024} or spin-polarized scanning tunneling microscopy—can directly verify the coexistence of the dissipative anomalous Hall response with the underlying layer-canting and non-coplanar helical spin texture.

Among realistic material platforms, parallel ($3R$-stacked) twisted homobilayers such as $\mathrm{MoTe}_2$ and $\mathrm{WSe}_2$ at small twist angles ($\theta \approx 2^\circ\text{--}4^\circ$) provide the most natural realization of our effective model \cite{wu_TopologicalInsulatorsTwisted_PhysRevLett_2019,pan_BandTopologyHubbard_PhysRevResearch_2020,anderson_ProgrammingCorrelatedMagnetic_Science_2023,cai_SignaturesFractionalQuantum_Nature_2023}. In these systems, the low-energy moiré bands map directly onto a buckled honeycomb Hubbard model in the strong-coupling limit ($U \gg t_1, t_2$), where competing exchange interactions naturally stabilize non-collinear $120^\circ$ and spiral spin textures \cite{pan_BandTopologyHubbard_PhysRevResearch_2020}. Dual electrostatic gating then directly tunes the layer potential asymmetry $D$ and carrier doping across the Chern metal phase boundaries.

In summary, we have investigated the emergence of topological metallic phases in strongly correlated moiré TMD bilayers hosting a spin-charge separated state coupled to a layer-polarized AFM spiral spin texture. Using the $su(2|1)$ coherent-state formalism, we derived an effective low-energy Hamiltonian for spinless holons propagating through a non-coplanar conical spinon background. We demonstrated that the interplay between the spiral pitch $\mathbf{Q}$, intralayer hopping $t_2$, and layer magnetizations $\mathcal{M}_{u,l}$ induces pronounced asymmetric band warping. While this warping closes the global indirect gap across most of the parameter space—confining true topological insulators to fine-tuned critical trajectories—a finite local direct gap persists, stabilizing broad regions of a topological Chern metal ($C = \pm 1$) with finite Fermi-level density of states. Our results establish a purely single-band strong-coupling mechanism for dissipative anomalous Hall transport ($R_{xy} \neq 0$ alongside finite $R_{xx}$) in dual-gated moiré materials.

\begin{acknowledgments}
K.K.K would like to thank Vikram Tripathi, Aleksander Razkov, Pavel Maksimov, Alvaro Ferraz for stimulating discussions and helpful comments on the manuscript. K.K.K acknowledges the financial support from the Foundation for the Advancement of Theoretical Physics and Mathematics ”Basis” for grant \# 23-1-4-63-1.
\end{acknowledgments}

\appendix
\section{Derivation of the Effective Holon Hamiltonian in the $su(2|1)$ Representation}
\label{app:su21-derivation}

In this Appendix, we provide the detailed derivation of the effective low-energy holon Hamiltonian [Eqs.~\eqref{eq:H_k_defn-doped}--\eqref{eq:defn-gamma-k-nnn}] starting from the interacting moiré-Hubbard model on the honeycomb superlattice [Eq.~\eqref{eq:ham-elec}].

\subsection{Hubbard Operator Projection and $su(2|1)$ Coherent States}

In the strong correlation regime ($U \gg t_1, t_2$), doubly occupied states $\ket{\uparrow\downarrow}$ on both sublattices are energetically prohibited and projected out of the low-energy Hilbert space using the Gutzwiller projection operator:
\begin{equation}
  \mathrm{P} = \prod_{i \in A} \left(1 - n_{i,A,\uparrow} n_{i,A,\downarrow}\right) \prod_{j \in B} \left(1 - n_{j,B,\uparrow} n_{j,B,\downarrow}\right).
\end{equation}
Acting on Eq.~\eqref{eq:ham-elec}, the on-site Coulomb $U$ term drops out, and the projected Hamiltonian expressed in terms of the sublattice-specific Hubbard projection operators $X_{i,\alpha}^{pq} = \ket{p}_{i,\alpha}\bra{q}_{i,\alpha}$ ($\alpha \in \{A, B\}$, $p,q \in \{\uparrow, \downarrow, 0\}$) reads:
\begin{widetext}
\begin{equation}
  \label{eq:app-ham-elec-X}
  \begin{aligned}
    H_{\text{proj}} = &-t_1 \sum_{\boldsymbol{\delta}_{nn}} \sum_{\sigma} \sum_{i \in A} \left( X_{i,A}^{\sigma 0} X_{i+\boldsymbol{\delta}_{nn},B}^{0\sigma} + \text{H.c.} \right) \\
                      &-t_2 \sum_{\boldsymbol{\delta}_{nnn}} \sum_{\sigma} \left( \sum_{i \in A} X_{i,A}^{\sigma 0} X_{i+\boldsymbol{\delta}_{nnn},A}^{0\sigma} + \sum_{j \in B} X_{j,B}^{\sigma 0} X_{j+\boldsymbol{\delta}_{nnn},B}^{0\sigma} \right) \\
                      &+ D \sum_{\sigma} \left( \sum_{i \in A} X_{i,A}^{\sigma\sigma} - \sum_{j \in B} X_{j,B}^{\sigma\sigma} \right) - \mu \sum_{\sigma} \left( \sum_{i \in A} X_{i,A}^{\sigma\sigma} + \sum_{j \in B} X_{j,B}^{\sigma\sigma} \right),
  \end{aligned}
\end{equation}
\end{widetext}
where $\boldsymbol{\delta}_{nn}$ ($n=1,2,3$) denote the three interlayer nearest-neighbor (NN) vectors connecting sublattices $A$ and $B$, and $\boldsymbol{\delta}_{nnn}$ denote the six intralayer next-nearest-neighbor (NNN) vectors connecting sites within the same sublattice.

To treat spin-charge fractionalization without imposing external slave-particle constraints or artificial gauge fields, we represent the local states on each sublattice via $su(2|1)$ coherent states \cite{perelomov_GeneralizedCoherentStates__1986,ferraz_EffectiveActionStrongly_NuclearPhysicsB_2011,kesharpu_MethodTreatStrongly_JPhysCondensMatter_2025,kesharpu_ProposalRealizingMajorana_PhysRevB_2024,kesharpu_ReentrantTopologicalOrder_PhysRevB_2025,kesharpu_TopologicalHallEffect_PhysRevB_2023,kesharpu_TopologicalHallEffect_PhysRevB_2024}:
\begin{equation}
  \label{eq:app-coherent-state}
  \ket{z_{i,\alpha}, \xi_{i,\alpha}}_{\alpha} = \frac{\ket{\uparrow}_{\alpha} + z_{i,\alpha}\ket{\downarrow}_{\alpha} + \xi_{i,\alpha}\ket{0}_{\alpha}}{\mathcal{N}_{i,\alpha}},
\end{equation}
where $z_{i,\alpha}$ are even Grassmann (complex) fields representing the localized spinon degrees of freedom ($[z_{i,A}, z_{j,B}] = 0$), and $\xi_{i,\alpha}$ are odd Grassmann fields representing the mobile, spinless holons ($\{\xi_{i,\alpha}, \xi_{j,\beta}\} = 0$). Because odd Grassmann variables are nilpotent ($\xi_{i,\alpha}^2 = \bar{\xi}_{i,\alpha}^2 = 0$), a Taylor expansion of the full state norm $\sqrt{1 + z_{i,\alpha}\bar{z}_{i,\alpha} + \xi_{i,\alpha}\bar{\xi}_{i,\alpha}}$ truncates at first order in $\xi_{i,\alpha}\bar{\xi}_{i,\alpha}$. After rescaling the Grassmann variables so that they satisfy canonical path-integral anti-commutation relations \cite{ferraz_EffectiveActionStrongly_NuclearPhysicsB_2011}, the effective normalization factors reduce purely to the spinon sector: $\mathcal{N}_{i,\alpha} = \sqrt{1 + z_{i,\alpha}\bar{z}_{i,\alpha}}$.

Projecting the Hubbard operators onto these local coherent states, $X_{i,\alpha}^{pq}(z,\xi) = \bra{z_{i,\alpha}, \xi_{i,\alpha}} X_{i,\alpha}^{pq} \ket{z_{i,\alpha}, \xi_{i,\alpha}}$, yields their classical superfield symbols \cite{kesharpu_MethodTreatStrongly_JPhysCondensMatter_2025,ferraz_EffectiveActionStrongly_NuclearPhysicsB_2011,kesharpu_ProposalRealizingMajorana_PhysRevB_2024}:
\begin{equation}
  \label{eq:app-X-symbols}
  \begin{aligned}
    X_{i,\alpha}^{\uparrow 0} &= \frac{\xi_{i,\alpha}}{\mathcal{N}_{i,\alpha}^2}, \qquad 
    X_{i,\alpha}^{\downarrow 0} = \frac{\bar{z}_{i,\alpha}\xi_{i,\alpha}}{\mathcal{N}_{i,\alpha}^2}, \\
    X_{i,\alpha}^{0\uparrow} &= \frac{\bar{\xi}_{i,\alpha}}{\mathcal{N}_{i,\alpha}^2}, \qquad 
    X_{i,\alpha}^{0\downarrow} = \frac{z_{i,\alpha}\bar{\xi}_{i,\alpha}}{\mathcal{N}_{i,\alpha}^2}, \\
    \sum_{\sigma} X_{i,\alpha}^{\sigma\sigma} &= 1 - X_{i,\alpha}^{00} = 1 - \frac{\xi_{i,\alpha}\bar{\xi}_{i,\alpha}}{\mathcal{N}_{i,\alpha}^2}.
  \end{aligned}
\end{equation}
Substituting Eq.~\eqref{eq:app-X-symbols} into the kinetic terms of Eq.~\eqref{eq:app-ham-elec-X} separates the holon hopping bilinears $\xi_i \bar{\xi}_j$ from the spinon kinematic overlap factors:
\begin{equation}
  \label{eq:app-ham-su21}
  \begin{aligned}
    H_{\text{kin}}(z,\xi) = &-t_1 \sum_{\boldsymbol{\delta}_{nn}} \sum_{i \in A} \left[ \frac{\xi_{i,A}\bar{\xi}_{i+\boldsymbol{\delta}_{nn},B}(1 + \bar{z}_{i,A} z_{i+\boldsymbol{\delta}_{nn},B})}{\mathcal{N}_{i,A}\mathcal{N}_{i+\boldsymbol{\delta}_{nn},B}} + \text{H.c.} \right] \\
                            &-t_2 \sum_{\boldsymbol{\delta}_{nnn}} \sum_{i \in A} \frac{\xi_{i,A}\bar{\xi}_{i+\boldsymbol{\delta}_{nnn},A}(1 + \bar{z}_{i,A} z_{i+\boldsymbol{\delta}_{nnn},A})}{\mathcal{N}_{i,A}\mathcal{N}_{i+\boldsymbol{\delta}_{nnn},A}} \\
                            &-t_2 \sum_{\boldsymbol{\delta}_{nnn}} \sum_{j \in B} \frac{\xi_{j,B}\bar{\xi}_{j+\boldsymbol{\delta}_{nnn},B}(1 + \bar{z}_{j,B} z_{j+\boldsymbol{\delta}_{nnn},B})}{\mathcal{N}_{j,B}\mathcal{N}_{j+\boldsymbol{\delta}_{nnn},B}}.
  \end{aligned}
\end{equation}

\subsection{Spinon Spiral Background, Gauge Transformation, and Momentum-Space Hamiltonian}

In the strong-coupling Mott-Hubbard regime ($U \gg t_1, t_2$), high-energy spinon amplitude fluctuations are suppressed by the Coulomb gap, and the low-energy magnetic background is governed by the layer-polarized conical spin spiral defined in Eqs.~\eqref{eq:spiral-spin-upper} and \eqref{eq:spiral-spin-lower}. In the singly occupied spinon sector, the local spin expectation values are related to the complex spinon field $z_{i,\alpha}$ ($\alpha \in \{A, B\}$) via the stereographic projection \cite{kesharpu_ReentrantTopologicalOrder_PhysRevB_2025,kesharpu_ProposalRealizingMajorana_PhysRevB_2024}:
\begin{equation}
  \label{eq:app-spin-ladder}
  S_{i,\alpha}^{+} = \frac{z_{i,\alpha}}{1 + |z_{i,\alpha}|^2}, \quad
  S_{i,\alpha}^{-} = \frac{\bar{z}_{i,\alpha}}{1 + |z_{i,\alpha}|^2}, \quad
  S_{i,\alpha}^{z} = \frac{1}{2}\left(\frac{1 - |z_{i,\alpha}|^2}{1 + |z_{i,\alpha}|^2}\right),
\end{equation}
where $S_{i,\alpha}^{x} = \frac{1}{2}(S_{i,\alpha}^{+} + S_{i,\alpha}^{-})$ and $S_{i,\alpha}^{y} = -\frac{\imath}{2}(S_{i,\alpha}^{+} - S_{i,\alpha}^{-})$. Inverting Eq.~\eqref{eq:app-spin-ladder} expresses the real and imaginary parts of the complex $c$-number field $z_{i,\alpha}$ directly in terms of the physical spin components:
\begin{equation}
  \label{eq:app-zi-cartesian}
  \operatorname{Re} z_{i,\alpha} = S_{i,\alpha}^{x}\left(1 + \frac{1 - 2S_{i,\alpha}^{z}}{1 + 2S_{i,\alpha}^{z}}\right), \quad
  \operatorname{Im} z_{i,\alpha} = S_{i,\alpha}^{y}\left(1 + \frac{1 - 2S_{i,\alpha}^{z}}{1 + 2S_{i,\alpha}^{z}}\right).
\end{equation}
Combining these components into polar form yields the exact one-to-one mapping:
\begin{equation}
  \label{eq:app-zi-polar}
  z_{i,\alpha} = \left(1 + \frac{1 - 2S_{i,\alpha}^{z}}{1 + 2S_{i,\alpha}^{z}}\right) \sqrt{\left(S_{i,\alpha}^{x}\right)^2 + \left(S_{i,\alpha}^{y}\right)^2} \exp\left[\imath \arctan\left(S_{i,\alpha}^{y} / S_{i,\alpha}^{x}\right)\right].
\end{equation}
Substituting our layer-polarized helical spin profiles [Eqs.~\eqref{eq:spiral-spin-upper} and \eqref{eq:spiral-spin-lower}] into Eq.~\eqref{eq:app-zi-polar} yields the real-space spinon fields in Eq.~\eqref{eq:zi-fields-opposite} of the main text, with uniform local normalization factors $\mathcal{N}_{i,A} = \sqrt{2/(1+\mathcal{M}_u)}$ and $\mathcal{N}_{j,B} = \sqrt{2/(1-\mathcal{M}_l)}$.

Evaluating the spinon overlap factors in Eq.~\eqref{eq:app-ham-su21} for these profiles reveals how the bond geometry controls the effective hopping phases. For the interlayer NN channels connecting sites $i \in A$ and $j = i + \boldsymbol{\delta}_{nn} \in B$, the opposite out-of-plane canting components give:
\begin{equation}
  \label{eq:app-zi-field-NN}
  \frac{1 + \bar{z}_{i,A} z_{j,B}}{\mathcal{N}_{i,A}\mathcal{N}_{j,B}} = \frac{1}{2}\left( \sqrt{\mathcal{M}_{u-l}} + \sqrt{\mathcal{M}_{l-u}}\,\mathrm{e}^{\imath \mathbf{Q}\cdot\boldsymbol{\delta}_{nn}} \right),
\end{equation}
where $\mathcal{M}_{u-l} = (1 + \mathcal{M}_u)(1 - \mathcal{M}_l)$ and $\mathcal{M}_{l-u} = (1 + \mathcal{M}_l)(1 - \mathcal{M}_u)$. Conversely, for the intralayer NNN hopping channels, both sites belong to the same sublattice, introducing a chiral sign inversion between the upper and lower layers:
\begin{equation}
  \label{eq:app-zi-field-NNN-A}
  \frac{1 + \bar{z}_{i,A} z_{i+\boldsymbol{\delta}_{nnn},A}}{\mathcal{N}_{i,A}\mathcal{N}_{i+\boldsymbol{\delta}_{nnn},A}} = \left( \cos\frac{\mathbf{Q}\cdot\boldsymbol{\delta}_{nnn}}{2} - \imath \mathcal{M}_u \sin\frac{\mathbf{Q}\cdot\boldsymbol{\delta}_{nnn}}{2} \right) \mathrm{e}^{\imath \mathbf{Q}\cdot\boldsymbol{\delta}_{nnn}/2},
\end{equation}
and for sublattice $B$:
\begin{equation}
  \label{eq:app-zi-field-NNN-B}
  \frac{1 + \bar{z}_{j,B} z_{j+\boldsymbol{\delta}_{nnn},B}}{\mathcal{N}_{j,B}\mathcal{N}_{j+\boldsymbol{\delta}_{nnn},B}} = \left( \cos\frac{\mathbf{Q}\cdot\boldsymbol{\delta}_{nnn}}{2} + \imath \mathcal{M}_l \sin\frac{\mathbf{Q}\cdot\boldsymbol{\delta}_{nnn}}{2} \right) \mathrm{e}^{\imath \mathbf{Q}\cdot\boldsymbol{\delta}_{nnn}/2}.
\end{equation}

Inserting Eqs.~\eqref{eq:app-zi-field-NN}--\eqref{eq:app-zi-field-NNN-B} into Eq.~\eqref{eq:app-ham-su21}, and performing the local $U(1)$ gauge transformation $\xi_{i,A(B)} \to \xi_{i,A(B)}\,\mathrm{e}^{\imath \mathbf{Q}\cdot\mathbf{r}_i / 2}$ to absorb the phase factors $\mathrm{e}^{\imath \mathbf{Q}\cdot\boldsymbol{\delta}_{nnn}/2}$, the real-space kinetic Hamiltonian reduces to:
\begin{widetext}
\begin{equation}
  \label{eq:app-ham-su21-real}
  \begin{aligned}
    H_{\text{kin}}(z,\xi) = &-t_1 \sum_{\boldsymbol{\delta}_{nn}} \sum_{i \in A} \left[ \xi_{i,A}\bar{\xi}_{i+\boldsymbol{\delta}_{nn},B} \left( \sqrt{\mathcal{M}_{u-l}}\,\mathrm{e}^{-\imath \frac{\mathbf{Q}\cdot\boldsymbol{\delta}_{nn}}{2}} + \sqrt{\mathcal{M}_{l-u}}\,\mathrm{e}^{\imath \frac{\mathbf{Q}\cdot\boldsymbol{\delta}_{nn}}{2}} \right) + \text{H.c.} \right] \\
                            &-t_2 \sum_{\boldsymbol{\delta}_{nnn}} \sum_{i \in A} \xi_{i,A}\bar{\xi}_{i+\boldsymbol{\delta}_{nnn},A} \left( \cos\frac{\mathbf{Q}\cdot\boldsymbol{\delta}_{nnn}}{2} - \imath \mathcal{M}_u \sin\frac{\mathbf{Q}\cdot\boldsymbol{\delta}_{nnn}}{2} \right) \\
                            &-t_2 \sum_{\boldsymbol{\delta}_{nnn}} \sum_{j \in B} \xi_{j,B}\bar{\xi}_{j+\boldsymbol{\delta}_{nnn},B} \left( \cos\frac{\mathbf{Q}\cdot\boldsymbol{\delta}_{nnn}}{2} + \imath \mathcal{M}_l \sin\frac{\mathbf{Q}\cdot\boldsymbol{\delta}_{nnn}}{2} \right).
  \end{aligned}
\end{equation}
\end{widetext}
Similarly, projecting the on-site potential terms of Eq.~\eqref{eq:app-ham-elec-X} via Eq.~\eqref{eq:app-X-symbols} and passing to the grand canonical ensemble for the holon sector incorporates the effective holon chemical potential term.


%

\end{document}